\documentclass[aps,preprint,preprintnumbers,nofootinbib,showpacs]{revtex4}
\usepackage{graphicx,color,amsmath}
\begin{document}

\preprint{%
\vbox{%
\hbox{hep-ph/0501265}
}}

\title{Splitting the split supersymmetry}

\author{Kingman Cheung}
\email[e-mail: ]{cheung@phys.nthu.edu.tw}
\affiliation{
Department of Physics and NCTS, National Tsing Hua University,
Hsinchu, Taiwan, R.O.C.}

\author{Cheng-Wei Chiang}
\email[e-mail: ]{chengwei@phy.ncu.edu.tw}
\affiliation{
Department of Physics, National Central University, Chungli, Taiwan 320,
R.O.C.}
\affiliation{
Institute of Physics, Academia Sinica, Taipei, Taiwan 115, R.O.C.}
\date{\today}

\begin{abstract}
  In split supersymmetry, the supersymmetric scalar particles are all very
  heavy, at least at the order of $10^9$ GeV, but the gauginos, Higgsinos, and
  one of the neutral Higgs bosons remain below a TeV.  Here we further split
  the split supersymmetry by taking the Higgsino mass parameter $\mu$ to be
  very large.  In this case, the $\mu$ problem is avoided and we keep the wino
  as a dark matter candidate.  A crude gauge coupling unification is still
  preserved.  Dark matter signals and collider phenomenology are discussed in
  this $\mu$-split SUSY scenario.  The most interesting dark matter signal is
  the annihilation into monochromatic photons.  In colliders, chargino-pair and
  the associated chargino-neutralino production cross sections have a certain
  ratio due to gauge couplings, and the chargino has long decays.
\end{abstract}
\preprint{}
\maketitle

\section{Introduction}

Supersymmetry (SUSY) is one of the most elegant solutions, if not the best, to
the gauge hierarchy problem.  The fine-tuning argument in the gauge hierarchy
problem requires SUSY particles at work at the TeV scale to stabilize the gap
between the electroweak scale and the grand unified theory (GUT) scale or the
Planck scale.  The most recent lower bound on the Higgs boson mass has been
raised to 114.4 GeV \cite{lep2}.  This in fact puts some stress on the soft
SUSY parameters, known as the little hierarchy problem.  Since the Higgs boson
receives radiative corrections dominated by the top squark loop, the mass bound
requires the top squark mass to be heavier than 500 GeV.  From the
renormalization-group (RG) equation of $M^2_{H_u}$, the magnitude of $M^2_{H_u}
\sim M^2_{\tilde{t}} \agt (500\; {\rm GeV})^2$.  Thus, the parameters in the
Higgs potential are fine-tuned at a level of a few percent in order to obtain a
Higgs boson mass of ${\cal O}(100)$ GeV.

Recently, Arkani-Hamed and Dimopoulos adopted a rather radical approach to SUSY
\cite{arkani}.  They essentially discarded the hierarchy problem by accepting
the fine-tuning solution to the Higgs boson mass.  They argued that since the
cosmological constant problem needs much more serious fine-tuning that one has
to live with, one may as well let go of the much less serious fine-tuning in
the gauge hierarchy problem.  The only criteria in setting up the scenario are
(i) the dark matter constraint imposed by the WMAP data \cite{wmap}:
$\Omega_{\rm DM} h^2 = 0.094 - 0.129$ ($2\sigma$ range), and (ii) the
gauge-coupling unification.
The scenario is coined as ``split SUSY'' \cite{giudice} with the spectrum
specified by the feature of the following distinct scales:
\begin{enumerate}
\item All the scalars, except for a CP-even Higgs boson, are very heavy.  One
  usually assumes a common mass scale for them at $\tilde{m} \sim 10^{9}$ GeV
  to $M_{\rm GUT}$.
\item The gaugino masses $M_i$ and the Higgsino mass parameter $\mu$ are
  comparatively much lighter and of the order of TeV in order to provide an
  acceptable dark matter candidate.
\end{enumerate}

In this work, we propose a further split in the split SUSY by raising the $\mu$
parameter to a large value which could be about the same as the sfermion mass
or the SUSY breaking scale.  We call it the $\mu$-split SUSY scenario.  In this
scenario, we do not encounter the notorious $\mu$ problem \cite{mu}.  At the
same time, our scenario can still achieve the gauge coupling unification and
provide a viable dark matter candidate.  The gauge coupling unification is
satisfied because the RG running of the gauge couplings is mainly determined by
the standard model (SM) particle and gaugino contributions.  Whether the
Higgsinos are very heavy has a milder effect.  The dark matter constraint
requires $M_2$ to be smaller than $M_1$; i.e., the dark matter is wino-like.

We summarize the differences between the split SUSY and our $\mu$-split SUSY
scenarios as follows.
\begin{enumerate}
\item The Higgsino mass parameter $\mu$ is raised to a very high scale in our
  scenario while in split SUSY it is at the electroweak scale.
\item The lightest supersymmetric particle (LSP) has to be the wino or gluino
  instead of the bino in our scenario because the bino would give a too large
  relic density, whereas the LSP can be the bino with $M_1 \sim \mu$ in split
  SUSY.
\item The wino dark matter can reach an interesting level of indirect
  detection, particularly the monochromatic photon signal, due to strong
  annihilation cross sections, and similarly for the anti-proton and positron
  detection.  On the other hand, the signal for direct detection is vanishingly
  small because of the absence of light squarks or Higgsino couplings.  In
  split SUSY both the direct and indirect detection signals are present,
  depending on the nature of the LSP.
\item In our $\mu$-split SUSY scenario, only chargino-pair production
  ($\widetilde{\chi}^+_1 \widetilde{\chi}^-_1$) and chargino-neutralino
  associated production ($\widetilde{\chi}^\pm_1 \widetilde{\chi}^0_1$) are
  possible at hadron colliders.  The cross sections are in a certain ratio in
  terms of gauge couplings.  Moreover, at $e^+ e^-$ colliders only the
  chargino-pair production is possible.  In split SUSY all pair production
  channels are possible.
\item In our $\mu$-split SUSY scenario, charginos can have long decays.  Since
  the mass difference between the chargino and the neutralino can be less than
  the pion mass, the chargino may travel more than a meter or so before it
  decays, and therefore producing ionized tracks in central silicon detectors.
  In split SUSY, the chargino decays promptly in general.
\end{enumerate}

The paper is organized as follows.  In the next section, we discuss a few
issues in raising the $\mu$ parameter.  We discuss the effects on gauge
coupling unification in Sec.~\ref{sec:coupling}, dark matter requirements in
Sec.~\ref{sec:dm}, and collider phenomenology in Sec.~\ref{sec:pheno}.  We
summarize our findings in Sec.~\ref{sec:summary}.

\section{Raising $\mu$ \label{sec:mu}}

In the minimal supersymmetric standard model (MSSM) a dimensionful
superpotential parameter $\mu$ is associated with the Higgs superfields.  A
natural choice for the value of this parameter should be either zero or the
scale of the ultra-violet (UV) theory, say, the Planck scale, GUT scale, or
SUSY breaking scale.  Nevertheless, phenomenological analyses give us a
different result.  Once the electroweak symmetry is broken, the weak scale,
characterized by the $Z$ boson mass, is given by the $\mu$ parameter and other
soft SUSY breaking parameters:
\begin{eqnarray}
\label{eq:mz}
\frac{M_Z^2}{2}
= \frac{M_{H_d}^2 - M_{H_u}^2 \tan^2\beta}{\tan^2\beta - 1} - \mu^2 ~.
\end{eqnarray}
Based upon the naturalness argument, $\mu$ along with other soft SUSY breaking
parameters are required to fall in a range near the weak scale.  This
discrepancy between scales of $\mu$ based on the two different naturalness
arguments is the so-called $\mu$ problem \cite{mu}.  It should be emphasized
that although the value of $\mu$ has a lower bound set by the chargino mass,
its upper bound purely comes from the naturalness requirement as outlined
above.

Since in the split SUSY model the fine-tuning of the light Higgs boson mass is
accepted, as is the case of the cosmological constant, we then do not need to
worry about the possibility of an unnatural cancellation between the two terms
on the right-hand side of Eq.~(\ref{eq:mz}).  Moreover, $\mu$, $M_{H_d}^2$, and
$M_{H_u}^2$ are now raised to more natural values, such as the SUSY breaking
scale, thus alleviating the $\mu$ problem.

Another issue is in the Higgs potential given by
\begin{equation}
\label{eq:vscalar}
 V_{\rm scalar}
 = (M_{H_u}^2 + \mu^2) |H_u|^2 + (M_{H_d}^2 + \mu^2) |H_d|^2 
   + \mu B (\epsilon_{ij} H_u^i H_d^j + {\rm h.c.}) + V_D ~,
\end{equation}
where $\epsilon_{12} = 1$, $V_D$ is the $D$-term contribution and much smaller
than the other terms in both the split SUSY and the $\mu$-split SUSY scenarios.
In order to have a light Higgs boson near the electroweak scale, a finely-tuned
relation among all three terms is required \cite{arkani}.  In the split SUSY,
since $\mu$ remains small, $B$ has to be extremely large ($\gg \tilde{m}$) such
that $\mu B$ is comparable to $M_{H_u}^2$ and $M_{H_d}^2$ (modulo an extra
small factor $\sim 1/\tan\beta$).  In the $\mu$-split SUSY, however, $\mu$ is
of the order $\tilde{m}$, the value of $B$ can just be of the same order such
that $\mu B \sim \tilde{m}^2$, the same order as $M_{H_u}^2$ and $M_{H_d}^2$.
In this sense, our $\mu$-split SUSY is better than split SUSY.

Moreover, one minimization condition of the Higgs potential,
Eq.~(\ref{eq:vscalar}), gives
\begin{eqnarray}
  \sin 2\beta = \frac{2 \mu B}{M_{H_u}^2 + M_{H_d}^2 + 2 \mu^2} ~.
\end{eqnarray}
As pointed out in Ref.~\cite{drees}, from the above expression split SUSY
predicts that $\tan\beta \sim {\cal O}(\tilde{m}^2 / M^2_{\rm weak})$, thereby
relating the two scales given the phenomenological constraint $0.5 \alt
\tan\beta \alt 100$.  This is because in split SUSY the light gaugino masses
are achieved by a softly broken $R$ symmetry.  However, such an $R$ symmetry
that allows a supersymmetric $\mu$ term forbids a nonvanishing $B$.  Therefore,
$|B| \sim {\cal O}(M_{\rm weak})$ and gives the above prediction of
$\tan\beta$.  In our scenario, we have already assumed $\mu \sim \tilde{m}$,
which is natural in the context of the $\mu$ problem, and the natural scale for
$B$ would be $M_{\rm SUSY} \sim \tilde{m}$, e.g., in gravity-mediated
models.\footnote {In our $\mu$-split SUSY scenario, we only need an $R$
  symmetry to forbid the gaugino masses.  Then, the $\mu$ parameter naturally
  takes on $\tilde{m}$ and so does the $B$ parameter.}  Thus, $\tan\beta \sim
{\cal O}(1)$, which fits easily within the phenomenological constraint $0.5
\alt \tan\beta \alt 100$.

\section{Gauge Coupling Unification \label{sec:coupling}}

The general form of the one-loop RG equations for the gauge couplings are given
by
\begin{equation}
\frac{1}{\alpha_i(M_X^2)} = \frac{1}{\alpha_i ( M_Z^2)} - 
 \frac{\beta_i}{4 \pi} \ln \left( \frac{M_X^2}{M_Z^2} \right ) ~,
\end{equation}
where $i = 1,2,3$ for the $SU(3)_C$, $SU(2)_L$, and $U(1)_Y$ gauge couplings,
respectively.  The differences among the SM, MSSM, split SUSY, and $\mu$-split
SUSY scenarios lie in the values of $\beta_i$'s:
\begin{eqnarray}
\mbox{SM}: && (\beta)_{\rm SM} = \left( \begin{array}{c}
                   0 \\   
                  - \frac{22}{3} \\
                  - 11 \end{array}         \right )
           +  \left( \begin{array}{c}
                   \frac{4}{3} \\   
                   \frac{4}{3} \\
                   \frac{4}{3}    \end{array} \right ) F 
           +  \left( \begin{array}{c}
                   \frac{1}{10} \\   
                   \frac{1}{6} \\
                   0            \end{array} \right ) N_H ~,
  \nonumber \\
\mbox{MSSM}: &&  (\beta)_{\rm MSSM} = \left( \begin{array}{c}
                    0 \\   
                  - 6 \\
                  - 9 \end{array}         \right )
           +  \left( \begin{array}{c}
                   2 \\   
                   2 \\
                   2    \end{array} \right ) F 
           +  \left( \begin{array}{c}
                   \frac{3}{10} \\   
                   \frac{1}{2} \\
                   0            \end{array} \right ) N_H ~,
  \nonumber \\
\mbox{Split-SUSY}: && (\beta)_{\rm split}|_{< \tilde{m}}= \left( \begin{array}{c}
                    0 \\   
                  - 6 \\
                  - 9 \end{array}         \right )
           +  \left( \begin{array}{c}
                   \frac{4}{3} \\   
                   \frac{4}{3} \\
                   \frac{4}{3}    \end{array} \right ) F 
           +  \left( \begin{array}{c}
                   \frac{5}{10} \\   
                   \frac{5}{6} \\
                   0            \end{array} \right ) ~,
  \nonumber \\
\mbox{$\mu$-split SUSY scenario}: && 
(\beta)_{\mu{\rm -split}}|_{< \tilde{m}}= \left( \begin{array}{c}
                    0 \\   
                  - 6 \\
                  - 9 \end{array}         \right )
           +  \left( \begin{array}{c}
                   \frac{4}{3} \\   
                   \frac{4}{3} \\
                   \frac{4}{3}    \end{array} \right ) F 
           +  \left( \begin{array}{c}
                   \frac{1}{10} \\   
                   \frac{1}{6} \\
                   0            \end{array} \right ) N_H ~,
  \nonumber
\end{eqnarray}
where $F=3$ is the number of generations of fermions or sfermions, and $N_H$ is
the number of Higgs doublets ($N_H=1$ in the SM, $N_H=2$ in the SUSY.)  In the
evolution of the gauge couplings in our scenario, we use
(i) the SM $\beta_i$'s from the weak scale ($M_Z$) to the scale of gaugino
masses, which we take a common value of 1 TeV;
(ii) the $\beta_i$'s in our $\mu$-split SUSY scenario from the gaugino mass
scale to the scalar mass scale ($\tilde{m}$), which we fix it at $10^9$ GeV;
and
(iii) the $\beta_i$'s for the MSSM from the scalar mass scale ($\tilde{m}$) to
the GUT scale.

\begin{figure}[t]
\centerline{
\includegraphics[height=8cm]{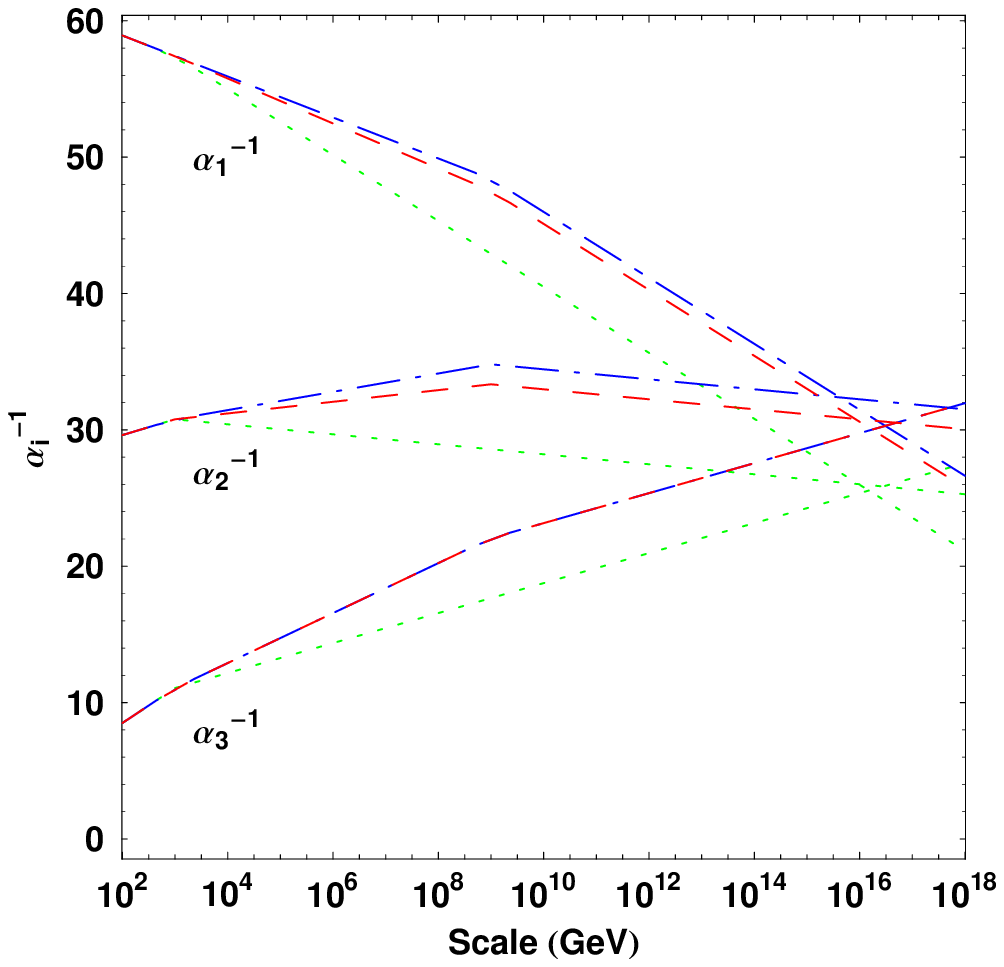}%
\hspace{0.5cm}
\includegraphics[height=8cm]{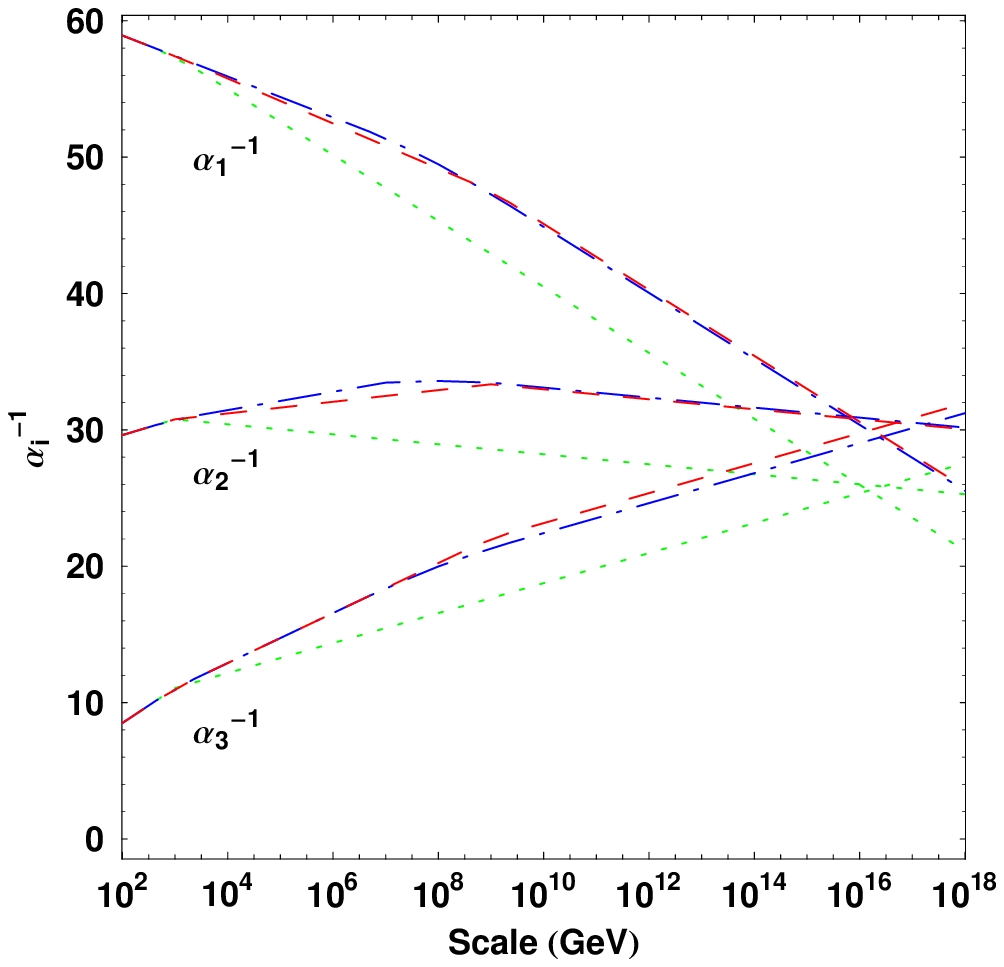}%
}
\centerline{(a) \hspace{8cm} (b)}
\caption{Gauge coupling unification in the MSSM (dotted green lines), the
  split SUSY (dashed red lines), and our $\mu$-split SUSY (dash-dotted blue
  lines) scenarios.  In plot (a), $\mu$ and the Higgsino masses are set the
  same as the sfermions at $10^9$ GeV.  Note that the curves for the strong
  coupling in split SUSY and our scenario coincide since the beta functions are
  identical.  In plot (b), $\mu$ and the Higgsino masses are fixed at $10^9$
  GeV, while the masses of the three generations of sfermions are separated
  into $10^7$ GeV, $10^8$ GeV, and $10^9$ GeV in our scenario as an example to
  illustrate possible threshold effects on the unification.  Following
  Ref.~\cite{arkani}, we take $\alpha_1^{-1}(M_Z) = 58.98$, $\alpha_2^{-1}(M_Z)
  = 29.57$, and $\alpha_3^{-1}(M_Z) = 8.40$ in making these plots.
\label{fig:couplings}
}
\end{figure}

In Fig.~\ref{fig:couplings}, we compare the scale dependence of the SM gauge
couplings in the MSSM, the split SUSY, and our $\mu$-split SUSY scenarios.  For
simplicity we take a universal value of 1 TeV for all the gaugino masses and a
value of $10^9$ GeV for all the SUSY scalars and Higgsino mass parameter $\mu$
in Fig.~\ref{fig:couplings}(a).  It is seen that our scenario shares a common
feature with the split SUSY; that is, the gauge couplings unify at ${\cal
  O}(10^{16})$ GeV and their unified value, $\alpha_{\rm GUT}$, is smaller than
that in the MSSM.  The imperfect unification can be used to account for the
discrepancy between the MSSM predicted strong coupling $\alpha_3^{\rm
  MSSM}(M_Z) = 0.130 \pm 0.004$, given the input of the experimental values of
$\alpha_{1,2}(M_Z)$, and the measured one $\alpha_3^{\rm exp}(M_Z) = 0.119 \pm
0.002$.  Although the triangular area enclosed by the three gauge coupling
curves in our scenario is larger than split SUSY, possible threshold effects
from sfermions may improve the situation.  This is illustrated in
Fig.~\ref{fig:couplings}(b), where we have separated the masses of the three
generations of sfermions into three different scales: $10^7$ GeV, $10^8$ GeV,
and $10^9$ GeV, respectively.  The two-loop contributions have minor effects on
these general behaviors.

%

\section{Dark Matter \label{sec:dm}}

\subsection{$M_1 < M_{2,3}$}

We are left with three gauginos in the TeV regime or less.  First of all, the
bino cannot be the lightest.  It is well-known that the bino annihilation cross
section is very small, so its relic density will be too large and overclose the
Universe if it is the LSP.  Therefore, this possibility is ruled out.
\footnote{ There is a slight complication \cite{stefano} when $\mu$ is not too
  large, e.g., 10 TeV, and $M_1$ is close to $M_2$ such that the LSP has a
  non-negligible fraction of wino and the LSP mass is close to the
  next-to-lightest neutralino and the light chargino.  In such a situation, the
  LSP can still annihilate efficiently to give the correct relic density.
  However, in our $\mu$-split-SUSY scenario, $\mu \sim 10^9$ GeV and thus $M_1$
  has to be extremely close to $M_2$ (they differ by $\alt 10^{-7}$) in order
  to have an effect.}

\subsection{$M_2 < M_{1,3}$}

In this case, the neutral wino is the LSP and has a large annihilation cross
section into $W$ pairs, $\widetilde{\chi}^0_1 \widetilde{\chi}^0_1 \to W^+
W^-$.  It constitutes a substantial fraction of the dark matter in the Universe
only if the wino mass is of order $2$ TeV.  The relic density for the wino dark
matter in the case of anomaly mediated SUSY breaking \cite{anom,anomaly} is
given by \cite{anomaly,aaron}
\begin{equation}
\Omega_{\widetilde{\chi}^0} h^2 \approx 0.05 \left( \frac{M_2}{\rm TeV} 
  \right )^2 \;.
\end{equation}
If the effect of coannihilation from the charged wino is included, the
coefficient in the above equation will be further reduced.  Therefore, if the
wino is the dominant dark matter component, its mass has to be of order 2 TeV.
However, there is also the possibility that the wino comes from other
non-thermal sources \cite{moroi}.

\subsection{$M_3 < M_{1,2}$}

One should also entertain the option that the gluino is the LSP, which has been
discussed in the literatures \cite{gunion,raby}.  The gluino can hadronize into
an $R$-hadron.  If the lightest $R$-hadron is electrically neutral and its mass
is of order $2-3$ TeV, it can also form a major component of the dark matter in
the Universe \cite{gunion}.  This is based on the assumption that the effective
annihilation of the $R$-hadron is due to the annihilation of its internal
gluino with the gluino from another $R$-hadron.  Since the annihilation of
gluinos is via strong interactions, the annihilation rate is typically large
and therefore requires a gluino mass of order $2-3$ TeV in order to be a dark
matter.  Also, the gluino LSP with a mass of 2 TeV or more is safe from the
search for strongly interacting massive particles in anomalously heavy nuclei
\cite{moha}.

\subsection{Other non-thermal sources}

Since the wino needs to be very heavy ($\sim 2$ TeV) to be the dominant dark
matter, the whole SUSY spectrum can only be marginally produced at the LHC.
The collider phenomenology will only be limited to a very small corner provided
by a $\sim 2-3$ TeV gluino.  The production rate is not high enough for a good
study.

On the other hand, there can be other non-thermal sources of dark matter.  For
example, in the context of anomaly-mediated SUSY breaking models
\cite{anom,anomaly} the LSP is usually the neutral wino.  For a relatively
light neutral wino it cannot be the dominant dark matter because of its large
annihilation cross sections.  However, an intriguing source of non-thermal wino
for compensation is the decay of moduli fields \cite{moroi}, which can produce
a sufficient amount of neutral winos.  In this case, even a light neutral wino
can constitute a major fraction of the dark matter.  There are also other
non-thermal sources of wino dark matter discussed in the literatures
\cite{nonthermal}.

\subsection{Dark matter detection}

Pure wino dark matter, as explained above, can come from both thermal and
non-thermal sources, with the latter source possibly being dominant.  The wino
dark matter has a very interesting signal for indirect detection in view of its
large annihilation cross sections, e.g., $\widetilde{\chi}^0_1
\widetilde{\chi}^0_1 \to W^+ W^-, \gamma \gamma, \gamma Z$. In particular, the
last two channels, though involving loop-suppressed cross sections, can give a
very clean signal of monochromatic photon lines.  If the resolution of the
photon detectors (either ground-based or satellite-based) is high enough, a
clean, sharp, unambiguous photon peak at hundreds of GeV should be observed
above the background.

Here we give an estimate on the photon flux in our $\mu$-split SUSY scenario.
We use the results given in Ref.~\cite{berg}.  In the limit of pure wino and
very heavy sfermions, only the $W^-$-$\widetilde{\chi}^+_1$ loop is important.
We obtain
\begin{equation}
v \sigma ( \widetilde{\chi}^0_1 \widetilde{\chi}^0_1 \to \gamma\gamma)
\simeq 14 \times 10^{-28} \;{\rm cm}^{3} {\rm s}^{-1} ~, \qquad
{\rm for }\;\; 
M_{\widetilde{\chi}^0_1} = 0.5-2 \;{\rm TeV} ~.
\end{equation}
Note that for comparison purposes, the value of $v\sigma$ for a pure Higgsino
case is about $1 \times 10^{-28} \;{\rm cm}^{3} {\rm s}^{-1}$ \cite{berg}.  The
photon flux as a result of this annihilation is given by \cite{berg2}
\begin{eqnarray}
\Phi_\gamma &\simeq& 1.87 \times 10^{-11} \, \left( \frac{N_\gamma v\sigma}
{10^{-29}\,{\rm cm}^3 {\rm s}^{-1}} \right ) \, \left( \frac{10\,{\rm GeV}}
{M_{\widetilde{\chi}^0_1}} \right )^2 \; J(\psi)\;
{\rm cm}^{-2} {\rm s}^{-1} {\rm sr}^{-1} \nonumber \\
&\simeq & 2 \times 10^{-10} \;{\rm cm}^{-2} {\rm s}^{-1} {\rm sr}^{-1}  \;,
\end{eqnarray}
where we have used $v\sigma=14\times 10^{-28}\,{\rm cm}^{3} {\rm s}^{-1}$,
${M_{\widetilde{\chi}^0_1}}=500$ GeV, $N_\gamma=2$, and $J(\psi=0)=100$ for the
photon flux coming from the Galactic Center.  The value of $J(\psi)$ depends on
the selected Galactic halo model.  It ranges from $O(10)$ to $O(1000)$
\cite{berg2}.  For a
typical Atmospheric Cerenkov Telescope (ACT) such as VERITAS \cite{veritas} and
HESS \cite{hess}, the angular coverage is about $\Delta \Omega =10^{-3}$ and
may reach a sensitivity at the level of $10^{-14} - 10^{-13} \, {\rm cm}^{-2}\,
{\rm s}^{-1}$ at the TeV scale.  Therefore, the signal of pure wino dark matter
annihilating into monochromatic photons is easily covered by the next
generation ACT experiments.

Since the wino annihilation into the $W^+ W^-$ pair is very effective, one can
also measure the excess in anti-protons and positrons \cite{anti}, 
which can be measured in
anti-matter search experiments, e.g., AMSII \cite{ams}.  We end this section by
noting that the direct search signal for our $\mu$-split SUSY scenario is very
difficult because of the absence of light squarks or the Higgsino components in
the lightest neutralino.

\section{Collider Phenomenology \label{sec:pheno}}

The collider phenomenology is mainly concerned with the production and
detection of neutralinos, charginos, and gluinos.  We restrict our discussions
below to cases with exact or approximate $R$-parity symmetry as follows.

\subsection{Neutralinos and Charginos}

Since the Higgsino mass parameter $\mu$ is very large, only the first two
neutralinos and the first chargino are light.  We will concentrate on their
phenomenology in this section.  Let us first examine their relevant couplings
to gauge bosons and the Higgs bosons.
\begin{itemize}
\item The $Z$-$\widetilde{\chi}^0_{1,2}$-$\widetilde{\chi}^0_{1,2}$ couplings
  only receive contributions from the Higgsino-Higgsino-gauge couplings.  In
  the limit of very large $\mu$ the Higgsino component of
  $\widetilde{\chi}^0_1$ and $\widetilde{\chi}^0_2$ are essentially zero.
  Therefore, they are zero.
\item The $H$-$\widetilde{\chi}^0_{1,2}$-$\widetilde{\chi}^0_{1,2}$ couplings
  have sources from the Higgs-Higgsino-gaugino terms.  Therefore, in the limit
  of large $\mu$, these couplings are also zero.
\item The \underline{$W^-$-$\widetilde{\chi}^0_{1,2}$-$\widetilde{\chi}^+_1$}
  couplings have sources from the Higgsino-Higgsino-gauge couplings and from
  the gaugino-gaugino-gauge couplings.  In the limit of large $\mu$, the former
  contribution goes to zero while the latter remains.  Therefore, the
  $W^-$-$\widetilde{\chi}^0_{1,2}$-$\widetilde{\chi}^+_1$ couplings contain
  only the gaugino-gaugino-gauge part.  Since only the wino component couples
  to the $W$ boson, thus only
  $W^-$-$\widetilde{\chi}^0_{1}$-$\widetilde{\chi}^+_1$ is nonzero if $M_2 <
  M_1$, and vice versa.
\item The $H^-$-$\widetilde{\chi}^0_{1,2}$-$\widetilde{\chi}^+_1$ couplings
  have sources from the Higgs-Higgsino-gaugino couplings. In the limit of large
  $\mu$, they do not contribute to
  $H^-$-$\widetilde{\chi}^0_{1,2}$-$\widetilde{\chi}^+_1$, which thus vanishes.
\item The \underline{
    $\gamma(Z)$-$\widetilde{\chi}^+_{1}$-$\widetilde{\chi}^-_1$} coupling has
  sources from the Higgsino-Higgsino-gauge couplings and from the
  gaugino-gaugino-gauge couplings.  In the limit of large $\mu$, the former
  contribution goes to zero while the latter remains.  Therefore, the
  $\gamma(Z)$-$\widetilde{\chi}^+_{1}$-$\widetilde{\chi}^-_1$ coupling contains
  only the gaugino-gaugino-gauge part.
\item The $H$-$\widetilde{\chi}^+_{1}$-$\widetilde{\chi}^-_1$ couplings have
  sources from the Higgs-Higgsino-gaugino couplings. In the limit of large
  $\mu$, they do not contribute to
  $H$-$\widetilde{\chi}^+_{1}$-$\widetilde{\chi}^-_1$, which thus vanishes.
\end{itemize}
The underlined are the only couplings that survive in the limit of large $\mu$
and large sfermion masses.  The phenomenology of the two light neutralinos and
the lightest chargino depends on the above nonzero couplings.

The parameter space of the SUSY spectrum relevant for phenomenology consists of
the bino ($M_1$), wino ($M_2$), and gluino ($M_3$) masses.  In the rest of the
paper, we only discuss the case of $M_2 < M_1$ because, as explain in the last
section, the bino LSP would overclose the Universe.

\begin{figure}[t!]
\centering
\includegraphics[width=5.5in]{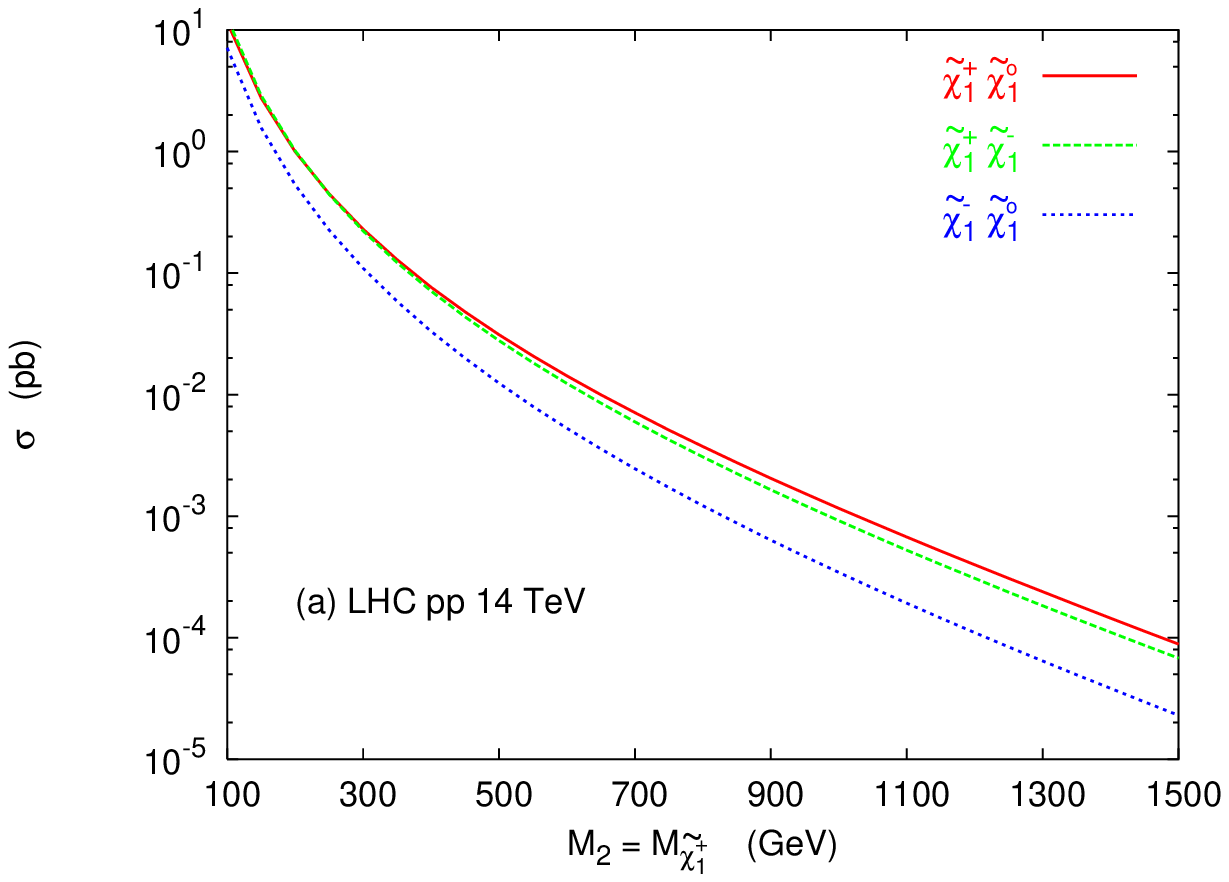}
\includegraphics[width=5.5in]{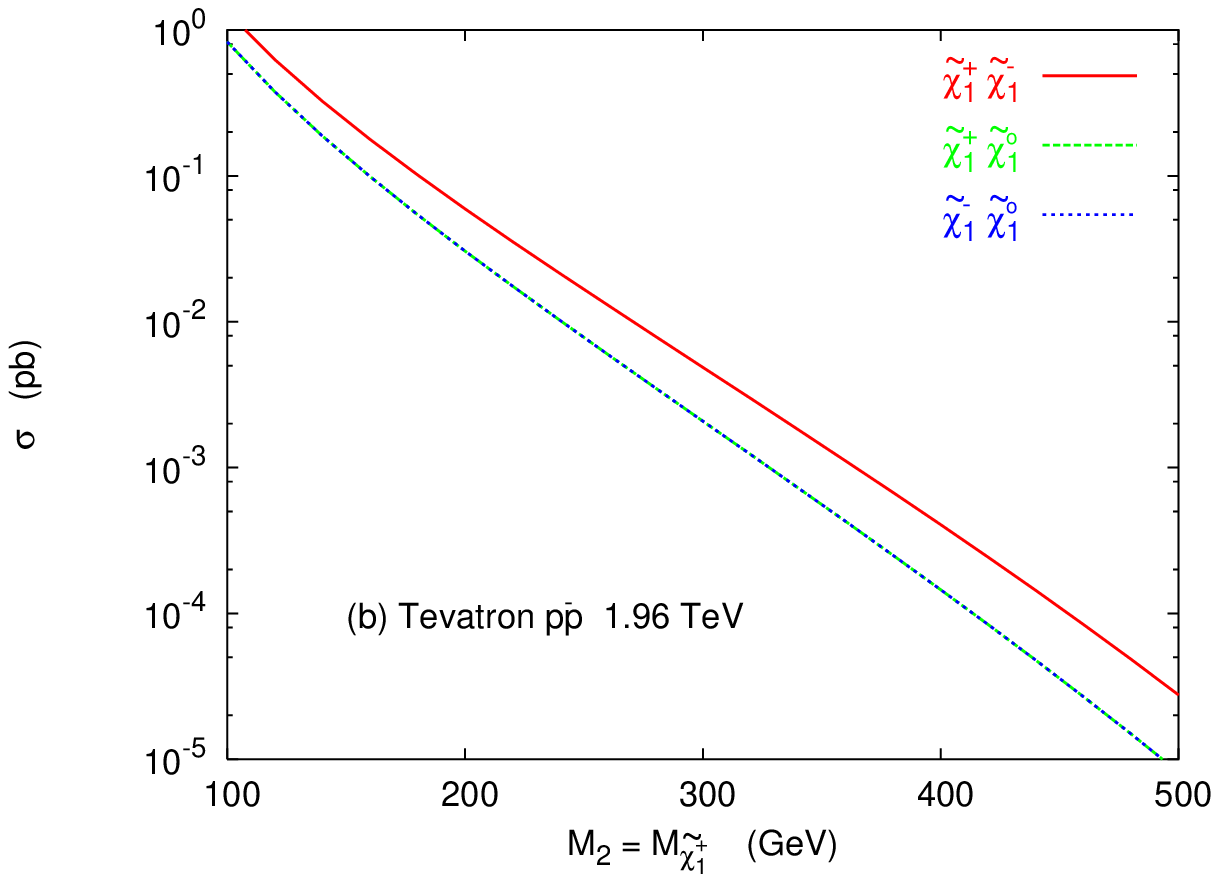}
\caption{\small \label{rate}
Production cross sections versus $M_2$ (the wino mass at the weak scale)
for the
$\widetilde{\chi}^+_1 \widetilde{\chi}^-_1$,
$\widetilde{\chi}^0_1 \widetilde{\chi}^+_1$, and
$\widetilde{\chi}^0_1 \widetilde{\chi}^-_1$ channels at (a) the LHC and 
(b) the Tevatron.  Note that at the Tevatron the production cross sections of 
$\widetilde{\chi}^0_1 \widetilde{\chi}^+_1$ and
$\widetilde{\chi}^0_1 \widetilde{\chi}^-_1$ are the same.
}
\end{figure}

As have discussed above, under the assumptions of large sfermion masses, large
$\mu$, and $M_2 < M_1$, the only sizable couplings to a pair of neutralinos and
charginos are
\[
\gamma (Z)\widetilde{\chi}^+_1 \widetilde{\chi}^-_1 \qquad
W^- \widetilde{\chi}^0_1 \widetilde{\chi}^+_1  \;.
\]
Note that the second lightest neutralino is almost a bino and thus has no
coupling to the $W$ boson.  In view of these couplings, the only noticeable
pair production channels at hadronic colliders are $\widetilde{\chi}^+_1
\widetilde{\chi}^-_1$ and $\widetilde{\chi}^0_1 \widetilde{\chi}^\pm_1$.  We
show the production cross sections of these channels versus $M_2$ (at the weak
scale) in Fig.~\ref{rate}.  Note that in the scenario with very large $\mu$ and
$M_2< M_1$ the lightest neutralino and chargino have masses very close to
$M_2$.  In fact, the conventional $U$ and $V$ matrices (that diagonalize the
chargino mass-squared matrix) are unit matrices.  Therefore, their masses are
simply equal to $M_2$ before radiative corrections are taken into account.
Radiative corrections lift the mass degeneracy and make the neutral wino
lighter than the charged wino \cite{radiative}.  Note also that the production
cross sections shown in Fig.~\ref{rate} are independent of $\tan\beta$.  This
very special scenario can then be checked by comparing gaugino-pair production
cross sections, shown in Fig. \ref{rate}, because they are simply given by the
gauge couplings.

In $e^+ e^-$ linear colliders, it is not possible to produce neutralino pairs
because of the absence of the Higgsino couplings.  Instead, only the chargino
pair production is possible.  This is another interesting check on our
$\mu$-split SUSY scenario.  In contrast, both neutralino-pair and chargino-pair
production are possible in the split SUSY scenario.

Another technical issue is to detect the chargino in either chargino-pair
production or the associated production with the lightest neutralino.  The
decay time of the chargino into the neutralino and a virtual $W$ boson depends
critically on the mass difference $\Delta M \equiv M_{\widetilde{\chi}^+_1} -
M_{\widetilde{\chi}^0_1}$.  The phenomenology in this case had been studied in
great details in Ref.~\cite{chen}.  We give highlights in the following
paragraph.

The partial width of $\widetilde{\chi}^+_1 \to \widetilde{\chi}^0_1 f \bar f'$
is given by \cite{chen}
\begin{eqnarray}
&& \Gamma (\widetilde{\chi}^+_1 \to \widetilde{\chi}^0_1 f \bar f') 
\nonumber \\
&& \qquad
= \frac{N_c(f) G_F^2}{(2\pi)^3} \biggr\{
 M_{\widetilde{\chi}^+_1} \left[ 
   \left( O^L_{11} \right)^2 +    \left( O^R_{11} \right)^2 \right]
\nonumber \\
&& \qquad\quad
\times \int_{( M_{\widetilde{\chi}^0_1} + m_f )^2}^{M^2_{\widetilde{\chi}^+_1
}} d q^2 \, \left( 1 - \frac{ M^2_{\widetilde{\chi}^0_1} + m^2_f} {q^2} 
           \right ) \,
   \left( 1- \frac{q^2}{ M^2_{ \widetilde{\chi}^+_1} } \right )^2 
   \sqrt{ \lambda( q^2,   M^2_{\widetilde{\chi}^0_1}, m^2_f ) }
                \nonumber \\
&& \qquad\quad
- 2  M_{\widetilde{\chi}^0_1} O^L_{11} O^R_{11}  \,
  \int_{m^2_f}^{ (M_{\widetilde{\chi}^+_1} -  M_{\widetilde{\chi}^0_1})^2}\;
 dq^2 \; \frac{q^2}{  M^2_{\widetilde{\chi}^+_1}} 
 \left( 1- \frac{ m^2_f}{q^2} \right )^2 
  \,    \sqrt{ \lambda(M^2_{\widetilde{\chi}^+_1},
 M^2_{\widetilde{\chi}^0_1}, q^2 ) } \biggl\} ~,
\label{decay}
\end{eqnarray}
where $(f,f')$ is, for example, $(u,d)$ or $(e,\nu_e)$, $N_c = 3 (1)$ if $f$ is
a quark (lepton), and $\lambda(a,b,c) = (a+b-c)^2 - 4 a b$.  The above formula
is valid for (i) leptonic decays, and (ii) hadronic decays when $\Delta M \agt
2$ GeV.  For hadronic decays with $\Delta M \alt 1-2$ GeV, one has to
explicitly sum over exclusive hadronic final states.  We have to include the
partial widths into one, two, and three pions. The explicit formulas can be
found in Ref.~\cite{chen}.  In our $\mu$-split SUSY case, $O_{11}^L = O_{11}^R
= 1$.  The detection of the chargino depends on the magnitude of $\Delta M$:

\begin{enumerate}
\item $\Delta M < m_\pi$.  In this case, the only available decay modes are
  $e^+ \nu_e$ and $\mu^+ \nu_\mu$.  We use Eq.~(\ref{decay}) to estimate the
  decay width to be ${\cal O}(10^{-7})$ eV or less.  Therefore, the chargino
  will travel a distance of the order of a meter or longer before decaying, and
  will leave a heavily-ionized track in the vertex detector.  The leptons
  coming out are too soft to be detectable.  This signal of ionized tracks is
  essentially SM background-free.
\item $m_\pi < \Delta M < 1$ GeV.  This is the most difficult regime, and very
  much depends on the design of the central detector.  The important criteria
  are the decay length $c\tau$ of the chargino and the momentum of the pion
  from the chargino decay.  The decay length $c\tau$ may be long enough to
  travel through a few layers of the silicon vertex detector.  For example, if
  $m_\pi < \Delta M < 190$ MeV the chargino will typically pass through at
  least two layers of silicon chips \cite{chen}.  Since the pion is derived
  from the chargino decay, it is a non-pointing pion.  That is, the backward
  extrapolation of the pion track does not lead to the interaction point.  The
  resolution on the impact parameter $b_{\rm res}$ depends on the momentum of
  the pion $p_\pi \sim \sqrt{ (\Delta M)^2 - m^2_\pi}$ in the chargino rest
  frame.  The higher the momentum is, the better the resolution $b_{\rm res}$
  will be \cite{chen}.  Thus, detecting the signal involves the combination of
  detecting a track left in only a few layers of the silicon and identifying a
  nonzero impact parameter of the pion coming out of the chargino.  A detailed
  simulation is beyond the scope of the present paper.
\item $\Delta M \agt 1-2$ GeV.  We can use Eq.~(\ref{decay}) to estimate the
  total decay width of the chargino.  The decay width is large enough that it
  decays promptly, producing soft leptons, pions, or jets, plus large missing
  energies.  The problem is on the softness of the leptons, pions, or jets.
  Experimentally, their detection is difficult.  Only if $\Delta M$ is
  sufficiently large to produce hard enough leptons or jets can the chargino
  decay be detected.  Otherwise, one has to rely on some other methods, such as
  $e^+ e^- \to \gamma \widetilde{\chi}^+_1 \widetilde{\chi}^-_1 \to \gamma +
  \not\!{E}_T$, a single photon plus large missing energy above the SM
  background $e^+ e^- \to \gamma \nu\bar \nu$ \cite{chen}.  Unfortunately, the
  signal rate is ${\cal O}(\alpha_{\rm em})$ smaller than the chargino-pair
  production.  Such a method of detecting the single photon plus missing energy
  is more difficult at hadronic colliders.
\end{enumerate}

In summary, the detection of the chargino is easier when $\Delta M$ is very
small ($< m_\pi$) or when $\Delta M$ is large ($>$ a few GeV).  The former
gives charged tracks in the silicon detector whereas the latter gives prompt
leptons or jets plus missing energies.  The intermediate range presents a
challenge to experiments.  The questions are how many layers of silicon that
the chargino can travel and how well the resolution the non-pointing pion can
be.

In the case of pure wino LSP, the lightest neutralino and chargino are
essentially degenerate in mass at the tree level.  The radiative corrections
can lift the mass degeneracy.  In our $\mu$-split SUSY scenario, the mass
difference is due to the radiative corrections of the gauge boson loops
\cite{nojiri}, given by
\begin{equation}
\delta m_{\rm rad} = \frac{\alpha_{\rm em} M_{\widetilde{\chi}^0_1}}
{4 \pi s_W^2} \, \left[
f\left ( \frac{m_W}{ M_{\widetilde{\chi}^0_1} } \right )
- c_W^2 \, f\left ( \frac{m_Z}{ M_{\widetilde{\chi}^0_1} } \right )
- s_W^2 \, f (0)
 \right ] \;,
\end{equation}
where $s_W$ and $c_W$ are the sine and cosine of the Weinberg angle, and $f(a)
= \int_0^1 dx 2(1+x) \log(x^2 + (1-x)a^2)$.  Numerically, we obtain $\delta
m_{\rm rad} \approx 170$ MeV for $M_{\widetilde{\chi}^0_1}= 0.5-2$ TeV.
Therefore, $\Delta M$ may fall into the very difficult regime.  However, there
may be higher order corrections or effects from other sources that can further
increase or decrease $\Delta M$.

Note that when $\tilde{m}$ and $\mu$ are of ${\cal O}(10^9)$ GeV or above, the
gluino will not decay within the detector, and the second lightest neutralino
$\widetilde{\chi}^0_2$ cannot be produced in the collisions.  Thus, there is no
production of the second lightest neutralino in colliders in such an extreme
scenario.  However, if $\tilde{m}$ is of order $10^6$ GeV or less, the gluino
can decay within the detector.  It will produce the second lightest neutralino,
which in turn decays into the lightest neutralino via an intermediate squark or
slepton. Therefore, the decay time is long.  We will not go further into this
low $\tilde{m}$ case.

\subsection{Gluino}

In our $\mu$-split SUSY scenario, the behavior of the gluino is the same as in
the split SUSY scenario.  Once produced, the gluino is stable inside the
detector or longer.  The signature of the gluino as the $R$-hadrons has been
studied in a number of papers \cite{raby,gunion,kilian,rizzo}.  Essentially,
once the gluino is produced it hadronizes into an $R$-hadron by combining with
light quarks or a gluon.  When the $R$-hadron traverses through the detector,
it will lose energy to the detector material, thus producing the signal.  The
detectability depends on the $R$-hadron spectrum and their electric charges.
In fact, it involves large uncertainties because the $R$-hadron spectrum is not
clearly known.  Also, the event rate has a large range depending on whether the
$R$-hadron is electrically charged \cite{yee}.  Further complication arises
from the fact that there can be frequent swappings between various states of
the $R$-hadron when it collides with nucleons of the detector material.
Another possibly clean signature was proposed to detect the gluino-gluino bound
state called the gluinonium \cite{yee}.  Since the gluino is stable, a pair of
gluinos can exchange gluons between themselves to form a bound state.  The
gluinonium can then annihilate into a pair of heavy top or bottom quarks,
experimentally forming a sharp peak in the invariant mass spectrum of $M_{t\bar
  t}$ or $M_{b\bar b}$, provided the background can be efficiently suppressed
\cite{yee}.

\section{Conclusions \label{sec:summary}}

In the present paper we have considered an even more radical scenario than
split SUSY.  We call it $\mu$-split SUSY, characterized by the assumption that
the $\mu$ parameter is raised together with the scalar sfermion to the high
SUSY breaking scale.  The Higgsino masses are lifted accordingly.  The main
motivation is to solve the $\mu$ problem.  We have investigated the effects on
gauge coupling unification, dark matter constraints, and collider
phenomenology.

We have found that the gauge coupling unification is slightly worse than the
split SUSY scenario mainly because of the change in the running of $\alpha_2$.
Nevertheless, the effect is mild.  On the other hand, there is a rather big
change in the dark matter requirement.  Since the LSP does not have the
Higgsino component any more, the LSP cannot be the bino because otherwise the
relic density would be too large.  The only sensible LSP is then the wino, with
the wino dark matter receiving contributions from both thermal and non-thermal
sources such that its mass can be less than one TeV.  The wino dark matter has
a large annihilation cross section, and the annihilation into two photons is
expected to give rise to a very clean and sharp monochromatic photon line.  The
flux is well above the sensitivity of the future ACT experiments.

The collider phenomenology is also quite different from the usual MSSM and
split SUSY.  The only possible gaugino production channels are the gluino pair,
chargino pair, and associated chargino-neutralino pair.  The behavior of the
gluino will be the same as in split SUSY, i.e., a long-lived gluino.  However,
the chargino-pair and the associated chargino-neutralino production channels
are very different.  The only production channels at hadron colliders are
$\widetilde{\chi}^+_1 \, \widetilde{\chi}^-_1$ and $\widetilde{\chi}^\pm_1
\widetilde{\chi}^0_1$, whereas at $e^+ e^-$ colliders only
$\widetilde{\chi}^+_1 \, \widetilde{\chi}^-_1$ is possible.  The production
cross sections are in a certain ratio depending on the mass $M_2$.  No such
behavior is seen in the split SUSY scenario.  Furthermore, in our $\mu$-split
SUSY scenario the chargino has a long decay, which again is very different from
the split SUSY.  The differences are already detailed in the Introduction.

\section*{Acknowledgments}

We thank Mihoko Nojiri and Tatsuo Kobayashi for discussions.  K.~C.\ would like
to thank YITP and KEK for their hospitality where part of the work was done.
C.-W.\ C. is grateful to the hospitality of National Center for Theoretical
Sciences, Taiwan, during his visit when part of the work was done.  This
research was supported in part by the National Science Council of Taiwan
R.~O.~C.\ under Grant Nos.\ NSC 93-2112-M-007-025- and NSC 93-2119-M-008-028-.


\end{document}